\documentclass{appolb}
\usepackage{graphicx}
\usepackage{color}

\begin{document}
\title{Time-dependent method for many-body problems and its 
application to nuclear resonant systems%
\thanks{Presented at the XXXV Mazurian Lakes Conference on Physics, 
Piaski, Poland, September 3-9, 2017}%
}
\author{Tomohiro Oishi$^{1, \dagger}$ and Lorenzo Fortunato$^1$ 
\address{$^1$Department of Physics and Astronomy ``Galileo Galilei'', University of Padova, 
and I.N.F.N. Sezione di Padova, Italy \\
$^{\dagger}$E-mail: toishi@pd.infn.it \linebreak}\\ 
}




\newcommand{\bi}[1]{\ensuremath{\boldsymbol{#1}}}
\newcommand{\unit}[1]{\ensuremath{\mathrm{#1}}}
\newcommand{\oprt}[1]{\ensuremath{\hat{\mathcal{#1}}}}
\newcommand{\abs}[1]{\ensuremath{\left| #1 \right|}}
\newcommand{\ket}[1]{\ensuremath{\left| #1 \right>}}
\newcommand{\Braket}[1]{\ensuremath{\left< #1 \right>}}

\def \beq{\begin{equation}}
\def \eeq{\end{equation}}
\def \beqa{\begin{eqnarray}}
\def \eeqa{\end{eqnarray}}
\def \Schr{Schr\"odinger }
\def \sway{\nonumber \\}

\def \bir{\bi{r}}
\def \ubir{\bar{\bi{r}}}
\def \bip{\bi{p}}
\def \ubip{\bar{\bi{r}}}

\maketitle
\begin{abstract}
The decay process of the \textcolor{black}{schematic one-dimensional three-body system} is considered. 
A time-dependent approach is used in combination with a one-dimensional 
three-body model, which is composed of a heavier core nucleus and 
two nucleons, 
with the aim of describing its evolution in two-nucleon emission. 
The process is calculated from the initial state, in which 
the three ingredient particles are confined. 
In this process, two different types of emission can be found: 
the earlier process includes the emission of spatially correlated 
two-nucleon pair, like a dinucleon, 
whereas, at a subsequent time, 
all the particles are separated from each other. 
The time-dependent method can be a suitable option to investigate 
the meta-stable and/or open-quantum systems, 
where the complicated many-body dynamics should necessarily be 
taken into account. 
\end{abstract}

\section{Introduction}
Quantum resonance or meta-stability is a basic concept to understand several 
dynamical processes in atomic nuclei. 
Those include, {\it e.g.} two-proton or two-neutron emission \cite{2009Gri,2012Pfu,2016Kondo}, 
tetra neutron \cite{2016Kisamori,2017Fossez}, and alpha-clustering resonant 
states ({\it c.f.} Hoyle state of $^{12}$C) \cite{1952Salp,1954Hoyle,2016Suno,2017Smith}. 
By investigating these processes, 
we expect to obtain fundamental information on nuclear interaction, 
multi-spin dynamics, and/or quantum tunneling effect in systems 
with many degrees of freedom. 

On the theoretical side, however, the description of these meta-stable 
systems has been a long-standing problem. 
The usual quantum mechanics for bound states should be extended to deal with 
the meta-stability and the multi-particle degrees of freedom on equal footing \cite{2009Gri,2016Suno,2014Myo}. 
For this purpose, we have developed a time-dependent three-body model for 
theoretical and computational approach \cite{1987Gurv,2004Gurv,2012Maru,2014Oishi}. 
This method can provide an intuitive way to discuss even the broad-resonance system, 
whose lifetime is considerably short, and thus the multi-particle dynamics should 
be taken into account.

In this work, we perform a toy-model calculation 
to investigate the broad-resonance state. 
We utilize the time-dependent method to describe the 
scattering emission from the three-body localized state. 
In contrast to the radioactive processes, 
it is not guaranteed that this emission process can be attributed 
to a single quasi-stationary state, 
but we have to take into account the contribution from all the possible components.

In the next section, we employ one-dimensional three-body model as our testing field 
for time-dependent calculation. 
Section \ref{sec:31} is devoted to present our results and discussions. 
Finally, we summarize this article in section \ref{sec:41}.

\begin{figure}[t]
  \centerline{%
  \includegraphics[width = 0.7\hsize]{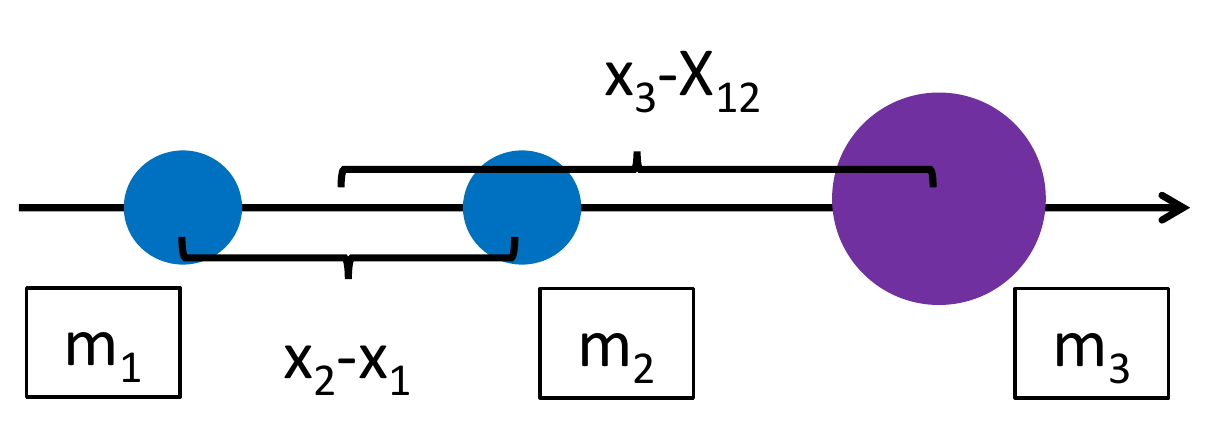}}%
  \caption{Three-body system in one dimension. $X_{12}=(x_1+x_2)/2$. } \label{fig:Ponch}
\end{figure}

\section{Model and Formalism} \label{sec:21}
\begin{figure}[t]
  \centerline{%
  \includegraphics[width = 0.7\hsize]{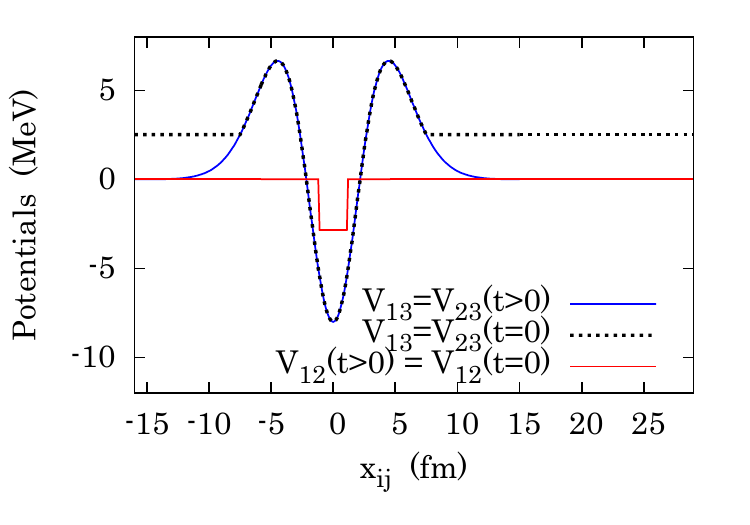}}%
  \caption{Two-body potentials as functions of the relative distances, $x_{ij}\equiv (x_i-x_j)$. } \label{fig:Vs}
\end{figure}

In this work, we give an example of the time-dependent (TD) calculation, 
implemented into a three-body system in one dimension (1D) \cite{2017FO}. 
The total Hamiltonian is 
\beq
 H_{tot} = \sum_{i=1}^{3} \frac{p_i^2}{2m_i} + V_{12}(\abs{x_1-x_2}) 
 + V_{23}(\abs{x_2-x_3}) + V_{13}(\abs{x_1-x_3}). 
\eeq
We employ the masses of particles defined as 
$m_1 = m_2 = 939$ MeV$/c^2$ and $m_3 = 16\cdot 939$ MeV$/c^2$. 
Namely, we assume a heavy core nucleus and two nucleons 
moving on the one-dimensional $x$-axis (see Fig.\ref{fig:Ponch}), 
mimicking the $^{18}$O nucleus but without pretending a realistic description. 
For the nucleon-nucleon subsystem, we employ a square-well 
attractive potential. 
That is, 
\beq
 V_{12}(x) =\left\{ \begin{array}{cc} 
                -2.84~~{\rm MeV}     & (\abs{x} \le 1.2~{\rm fm}), \\
                 0 \phantom{00000}   & (\abs{x} > 1.2~{\rm fm}). \end{array} \right.
\eeq
For the core-nucleon channel, on the other hand, 
\beq
 V_{13}(x)=V_{23}(x)= 
    V_r \exp \left( -\frac{x^2}{d^2_r} \right) 
  + V_a \exp \left( -\frac{x^2}{d^2_a} \right),
\eeq
where $d_r=5.04$ fm, $d_a=3.15$ fm, $V_r=24$ MeV and $V_a = -32$ MeV. 
These potentials are shown in Fig.\ref{fig:Vs}. 
\textcolor{black}{The bump in the core-nucleon potential 
can be associated with the centrifugal barrier in realistic nuclei. 
Note that, in this work, we focus on the 
broad-resonance state. 
For this purpose, the two-body potentials are fixed shallower than the 
usual potentials in the three-dimensional calculations. 
Also, instead of the Woods-Saxon type, we employ the Gaussian potential, 
which enables us to utilize the analytic formula to obtain the 
matrix elements with the harmonic oscillator (HO) basis 
employed in the next subsection. }

\subsection{Coordinates and Basis}
In order to solve the eigen-states of $H_{3b}$, first we employ the mass-scaled 
Jacobi coordinates (MSJC) \cite{1960Del,2002Aquil}. 
Using the common-relative mass, $\mu \equiv \sqrt{\prod_{i=1}^{3} m_i / \sum_{i=1}^{3}m_i}$, 
those are defined as, 
\beq
 \xi_1= \sqrt{\frac{\mu_1}{\mu}}(x_2-x_1), ~~~\xi_2= \sqrt{\frac{\mu_2}{\mu}} \left( x_3-\frac{x_2+x_1}{2} \right), 
\eeq
and $\xi_3 \equiv (m_1 x_1 + m_2 x_2 + m_3 x_3)/ \sum_{i=1}^{3}m_i$, 
which is the center-of-mass coordinate. 
Partial relative masses are defined as 
$\mu_k \equiv m_{k+1} \sum_{i=1}^k m_i / \sum_{j=1}^{k+1}m_j$, for 
$k=1$ and $2$. 
With MSJC, the total Hamiltonian reads 
\beqa
 H_{tot} &=& T_{CM} + \frac{\pi^2_1}{2\mu} + \frac{\pi^2_2}{2\mu} + V_{12}+V_{23}+V_{13}, \nonumber \\
 T_{CM} &=& \frac{\pi^2_3}{2(m_1+m_2+m_3)}, 
\eeqa
where $\left\{ \pi_i \right\}$ are the conjugate momenta to $\left\{ \xi_i \right\}$. 
In the following, we neglect the center-of-mass motion, $T_{CM}$. 
We diagonalize the remaining Hamiltonian, $H_{3b}=H_{tot}-T_{CM}$, 
by calculating its matrix elements, $\Braket{\Psi_{cd}| H_{3b} |\Psi_{ab}}$, 
within the harmonic oscillator (HO) basis: 
\beq
 \Psi_{ab}(\xi_1,\xi_2) = \psi_a(\xi_1) \psi_b(\xi_2), 
\eeq
where $a$ and $b$ are non-negative integers. 
Notice that $\psi_n$ is the HO wave function 
corresponding either to the relative motion of particles $1$ and $2$, 
or the motion of particle $3$ with respect to the center-of-mass between $1$ and $2$, 
with HO-energy, $(n+1/2) \hbar \omega$. 
Our model space is truncated as $a,b \le 15$ with $\hbar \omega = 0.4$ MeV. 
This value is chosen in a range that insure the convergence. 

In this article, we assume that two nucleons have the spin-singlet 
configuration: $\ket{S_{12}=0}=(\ket{\uparrow \downarrow}-\ket{\downarrow \uparrow})/\sqrt{2}$. 
Thus, the spatial part should be symmetric against the exchange between particles $1$ and $2$. 
It means that only $\left\{ \psi_a (\xi_1) \right\}$ with even $a$ can be included in our basis. 

Within the chosen MSJC scheme, 
the matrix elements of $V_{12}$ are diagonal, 
whereas $V_{23}$ and $V_{13}$ yield non-diagonal components. 
For computation of these non-diagonal elements, we utilized a 
kinetic rotation technique, whose details can be found in Ref. \cite{2017FO}. 
Then, all the eigen-states, $H_{3b}\ket{\Phi_M}=E_M\ket{\Phi_M}$, 
can be solved by diagonalization: $\ket{\Phi_M}=\sum_{ab} c_{M,ab} \ket{\Psi_{ab}}$.

\subsection{Initial State for Time Evolution}
We employ the confining potential method for time-evolution. 
This method has provided a good approximation for quantum 
meta-stable phenomena especially in nuclear physics \cite{1987Gurv,2004Gurv,2012Maru}. 
For the confining potential, $V_{13}^{(c)}=V_{23}^{(c)}$ at $t=0$ fm$/c$, 
we fix the wall potential from $\abs{x_i-x_j}\ge 7.5$ fm. 
On the other hand, $V_{12}$ between the light two particles is unchanged. 
See Fig.\ref{fig:Vs} for visual plots of these potentials. 
Our initial state, $\ket{\Upsilon(t=0)}$, is solved by diagonalizing 
the confining Hamiltonian including $V_{13}^{(c)}$ and $V_{23}^{(c)}$. 
It is also worthwhile to note that the initial state can be expanded on the 
eigen-states of the true Hamiltonian: 
\beq
 \ket{\Upsilon(t=0)} = \sum_M d_{M}(0) \ket{\Phi_{M}}. \label{eq:Rill}
\eeq
For this initial state, 
\textcolor{black}{ after the subtraction of the center-of-mass motion, 
the expectation value of the relative Hamiltonian is 
given as $\Braket{\Upsilon(0)|H_{3b}|\Upsilon(0)}=0.91$ MeV. 
This is equivalent to the energy release (Q-value) carried out 
by the emitted particles. }

\begin{figure}[t] \begin{center}
  \centerline{%
  \includegraphics[width = 0.65\hsize]{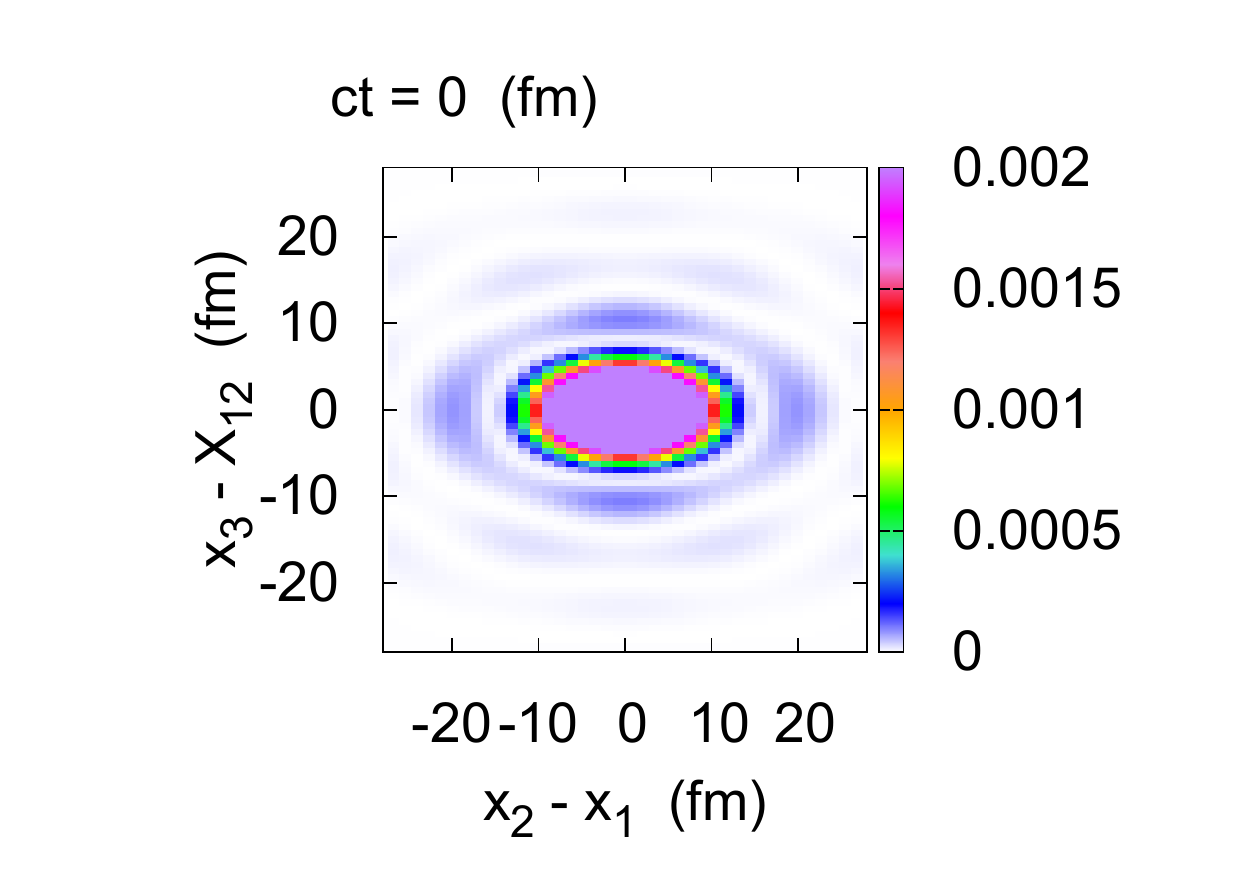}
  \hspace{-2.5cm}
  \includegraphics[width = 0.65\hsize]{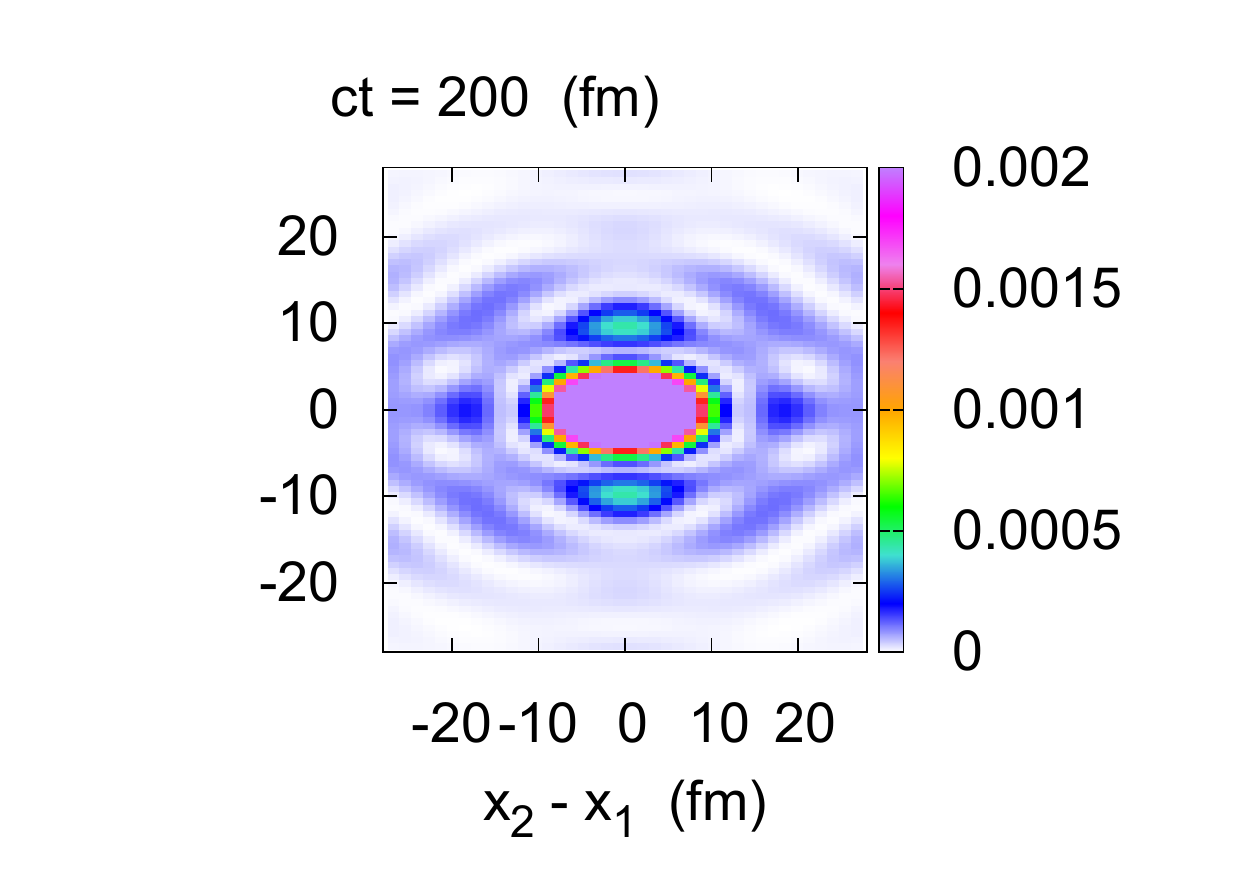} }%
  \centerline{%
  \includegraphics[width = 0.65\hsize]{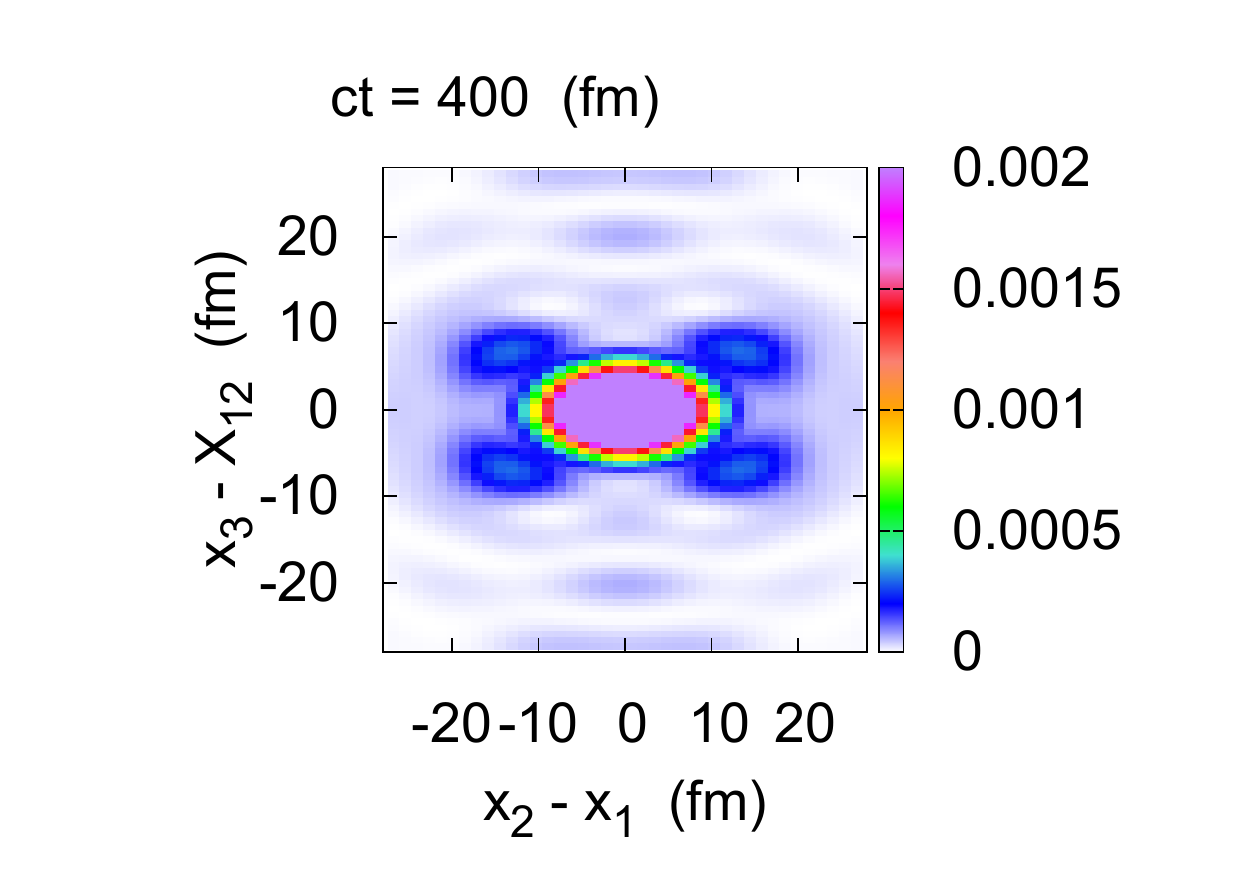}
  \hspace{-2.5cm}
  \includegraphics[width = 0.65\hsize]{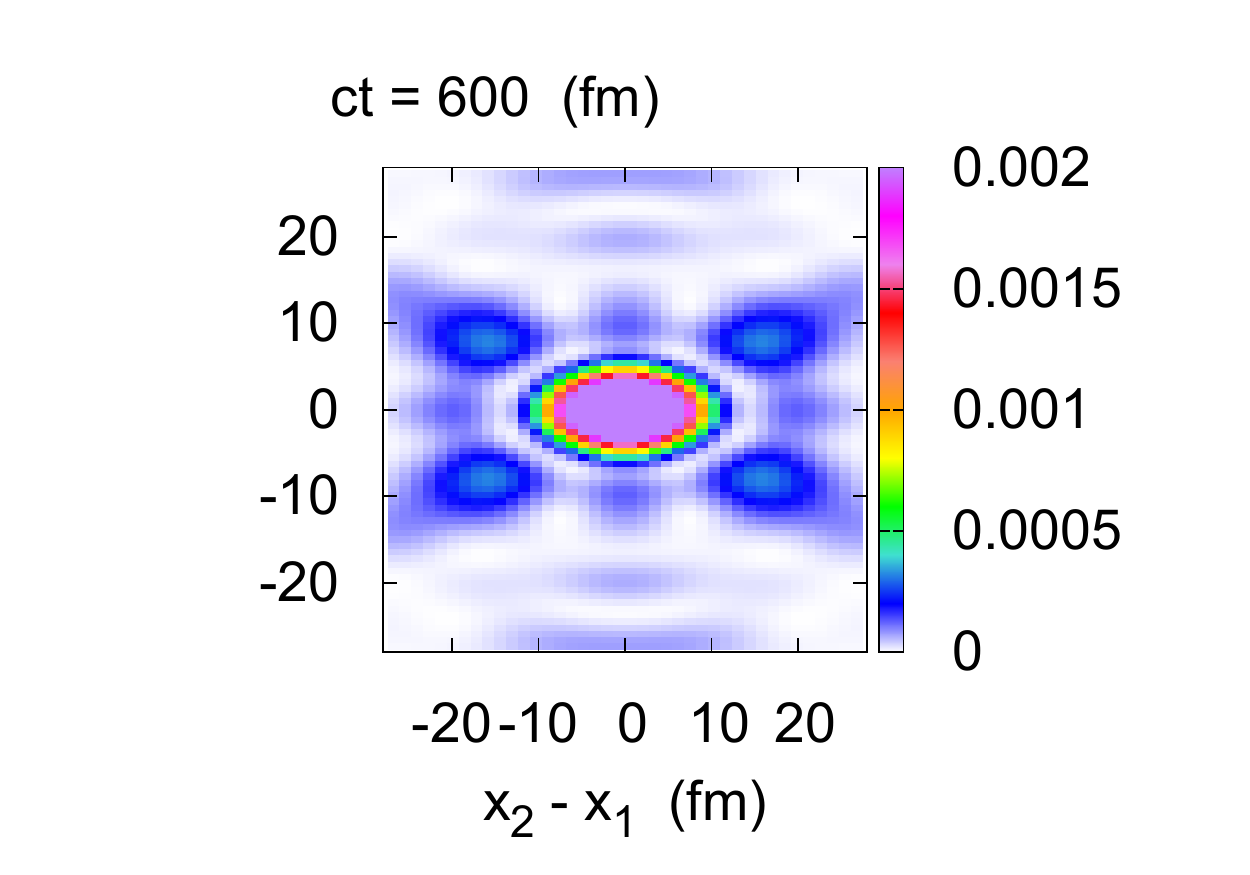}}%
  \caption{Density distribution, $\rho(t)=\abs{\Upsilon(t)}^2$, for 
$ct=0,200,400$ and $600$ fm. 
These are plotted as functions of $x_1-x_2$ and $x_3-X_{12}$, 
where $X_{12}$ is the center-of-mass between the 1st and 2nd particles. }
  \label{fig:imag_01}
\end{center} \end{figure}

\section{Result and Discussion} \label{sec:31} 
In the first panel of Fig.\ref{fig:imag_01}, we plot the density distribution 
of the initial state: $\rho(t=0)=\abs{\Upsilon(t=0;\xi_1,\xi_2)}^2$. 
As expected, the three ingredient particles are spatially localized at $t=0$. 

\begin{figure}[tb]
  \centerline{%
  \includegraphics[width = 0.5\hsize]{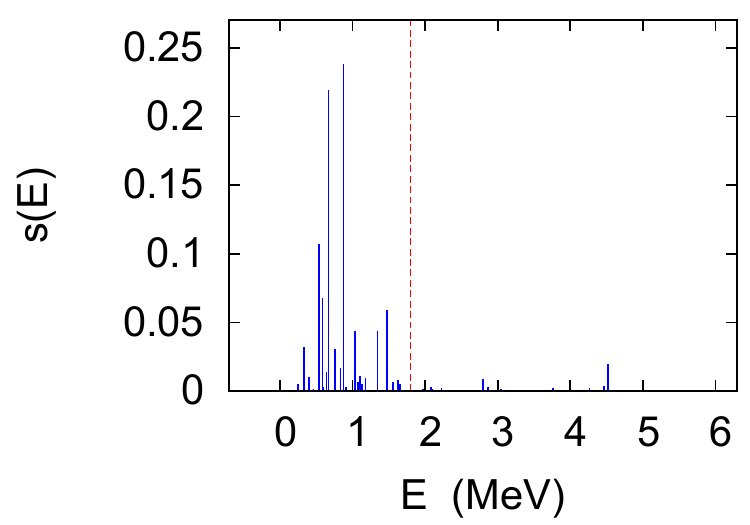}
  \includegraphics[width = 0.5\hsize]{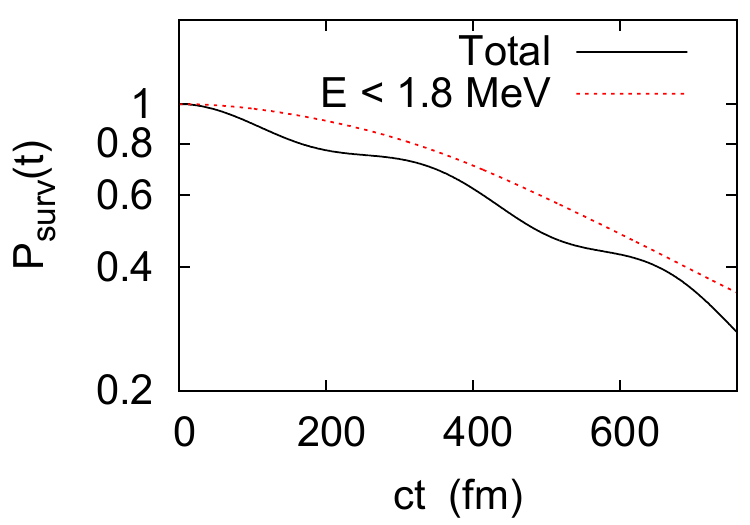}}%
  \caption{(Left panel) Energy spectrum of the emission state, $\ket{\Upsilon(t)}$. 
$E_T=1.8$ MeV is indicated by the dashed line. 
(Right panel) Survival probability. }
  \label{fig:Edist}
\end{figure}

\subsection{Time-dependent Emission}
From Eq.(\ref{eq:Rill}), time-evolution via $H_{3b}$ can be calculated as 
\beqa
 && \ket{\Upsilon (t)} \equiv  \exp \left[ -it \frac{H_{3b}}{\hbar} \right] \ket{\Upsilon(t=0)} 
 = \sum_{M} d_M(t) \ket{\Phi_M}, ~~~{\rm where} \nonumber \\
 && d_M(t) = e^{-itE_M/\hbar} d_M(0). 
\eeqa
The time-evolution of the density distribution is shown in Fig. \ref{fig:imag_01}. 
That is, 
\beq
 \rho(t; \xi_1,\xi_2) = \abs{\Upsilon(t; \xi_1,\xi_2)}^2. 
\eeq
Notice that the energy distribution is invariant during the time-evolution: 
$s(E_M) \equiv \abs{d_M(0)}^2 = \abs{d_M(t)}^2$. 
In Fig.\ref{fig:Edist}, we plot the energy distribution. 
From this result, we can find that 
the state of interest, $\ket{\Upsilon(t)}$, can be mostly attributed 
to the low-lying components with continuum energies up to $E\le 1.8$ MeV.

In Fig.\ref{fig:imag_01}, 
at $ct=200$ fm, the emission process proceeds mainly 
with $x_2=x_1$ and $x_3-X_{12}= \pm 10$ fm, 
where $X_{12}$ indicates $(x_1+x_2)/2$. 
This earlier process means that the two light particles, 
$m_1$ and $m_2$, are spatially correlated and emitted as a pair from the core. 
Namely, we observe a dinucleon emission in 1D space \cite{2012Maru}. 

After $ct\ge 400$ fm as shown in Fig.\ref{fig:imag_01}, 
on the other hand, the process shows a 
different pattern with $\abs{x_2-x_1} \simeq 15$ fm and 
$\abs{x_3-X_{12}} \simeq 10$ fm. 
In this process, the two light particles are not localized anymore, 
and three particles move away from each other. 
Thus, the total emission should be a superposition of the primary 
dinucleon emission and the secondary separated emission. 
\textcolor{black}{This superposition is quite in contrast to Ref.\cite{2012Maru}, 
where only the dinucleon emission is dominant with the pairing force. }
In such a way, 
our time-dependent method can provide a direct and intuitive 
solution to describe this complex quantum dynamics.

\subsection{Survival Probability}
First we define the decay state, $\ket{\Upsilon_d(t)}$, such as 
\beq
 \ket{\Upsilon_d(t)} \equiv \ket{\Upsilon(t)} - \beta(t) \ket{\Upsilon(0)}
 = \sum_M y_M (t) \ket{\Phi_M}, 
\eeq
where $\beta(t) \equiv \Braket{\Upsilon(0) | \Upsilon(t)}$ and 
$y_M(t) = d_M(t)-\beta(t)d_M(0)$. 
Notice that $\Braket{\Upsilon(0) |\Upsilon_d(t)}=0$. 
Also, the decay probability can be formulated as 
$P_{decay}(t) \equiv \Braket{\Upsilon_d(t)| \Upsilon_d(t)} = 1-P_{surv}(t)$, 
where $P_{surv}(t)$ is the so-called survival probability. 
That is, 
\beq
 P_{surv}(t) = \abs{\beta(t)}^2 = \abs{\Braket{\Upsilon(0) | \Upsilon(t) }}^2. 
\eeq
In the second panel of Fig.\ref{fig:Edist}, the survival probability 
is plotted in logarithmic scale: 
there is an oscillatory decay along time-evolution. 
Thus, this process is not alike the radioactive emission, 
since the exponential decay-rule is hardly observed. 

Indeed, the process can be interpreted as a superposition 
of the well-converged exponential decay and the 
fluctuation due to high-energy components. 
To confirm this, remembering that $P_{surv}(t)=1-P_{decay}(t)$, 
we decompose the decay probability into the low- and high-energy components 
by fixing the border of $E_T=1.8$ MeV. 
That is, 
\beqa
 P_{decay}(t) &=& \sum_{E_M < E_T} \abs{y_M(t)}^2 + \sum_{E_M \ge E_T} \abs{y_M(t)}^2 \nonumber \\
 &\equiv & P_{decay}(t;E<E_T) + P_{decay}(t;E \ge E_T). 
\eeqa
Then, in Fig.\ref{fig:Edist}, we plot the low-energy component 
of the survival probability: $P_{surv}(t;E<E_T)=1-P_{decay}(t;E<E_T)$. 
Consequently, it shows a smooth pattern and acquires an exponentially 
decaying form after $ct \ge 500$ fm. 
In this exponential decay, $P_{surv}(ct\ge 500~{\rm fm};E<E_T) \simeq \exp (-t\Gamma/ \hbar)$, 
where the decay-width is approximated as $\Gamma \simeq 0.41$ MeV in our calculation. 
Notice that this decay-width value is similar to the empirical values 
observed in several light one- and two-proton emitters \cite{2009Gri,2012Pfu}.

\section{Summary} \label{sec:41}
We have performed the time-dependent analysis of the emission 
process in the 1D three-body system. 
By monitoring the time-evolution from the initially confined state, 
we confirmed that two different types of emission are taking place: 
the earlier dinucleon emission, and the secondary separated emission. 
It is shown that, even for such a superposition of different 
processes, our time-dependent calculation can be a suitable 
tool to understand its multi-particle dynamics with an 
intuitive procedure. 
By analyzing the survival probability, we have also found that 
this process can be interpreted mainly as the exponential 
decay with $E<1.8$ MeV plus the higher energy fluctuation. 
\textcolor{black}{Further investigation of the origin of 
this fluctuation is an important future task. 
This investigation can lead to a deeper knowledge of the nuclear 
meta-stable systems, {\it e.g.} tetra neutron, }
whose measured decay-width is considerably wide and 
hardly allows us to infer an exponential-decay behavior \cite{2016Kisamori}. 
Our extension of the time-dependent method applied to these realistic 
3D nuclear systems is in progress now. 

This work is financially supported by the P.R.A.T. 2015 project 
{\it IN:Theory} in the University of Padova (Project Code: CPDA154713).




\end{document}